\documentclass{article}

\usepackage{amssymb,amsmath}
\usepackage{fullpage}
\usepackage{graphicx}
\usepackage{amsmath}		
\usepackage[margin=1.0in]{geometry}
\usepackage{setspace}
\usepackage{color}
\usepackage{fancyhdr}
\usepackage{collcell}
\usepackage{datatool}
\usepackage{environ}
\usepackage{latexsym}
\usepackage{amssymb}
\usepackage{epsfig,amsmath,graphics}
\usepackage{epstopdf}
\usepackage{verbatim}
\usepackage{wasysym}
\usepackage{feynmp-auto}
\usepackage{authblk}
\usepackage{xcolor}
\usepackage{enumitem}
\usepackage[utf8]{inputenc}
\usepackage{slashed}
\usepackage{cite}

\usepackage{empheq}

    \newlength\fsep
    \newsavebox\widebox

\usepackage[skip=0pt]{caption}

\def\TM#1{{\bf  \textcolor{blue}{[TM: {#1}]}}}

\begin{document}
\title{Time Evolution in Quantum Cosmology}
\author{Anne-Katherine Burns}
\affil{\small Department of Physics and Astronomy, University of California, Irvine, CA 92697 USA}
\author{David E.~Kaplan}
\affil{\small Department of Physics \& Astronomy, The Johns Hopkins University, Baltimore, MD  21218, USA}
\author{Tom Melia}
\affil{\small Kavli IPMU (WPI), UTIAS, The University of Tokyo, Kashiwa, Chiba 277-8583, Japan}
\author{Surjeet Rajendran}
\affil{\small Department of Physics \& Astronomy, The Johns Hopkins University, Baltimore, MD  21218, USA}

\date{\today}
\maketitle

\begin{abstract}

The quantum description of time evolution in non-linear gravitational systems such as cosmological space-times is not well understood. We show, in the simplified setting of mini-superspace, that time evolution of this system can be obtained using a gauge fixed path integral, as long as one does not integrate over proper time.  
Using this gauge fixed action we can construct a Hamiltonian in the coherent - or classical - state basis. We show that by construction the coherent states satisfy the classical dynamical equations of General Relativity. They do not satisfy the Hamiltonian constraint. A consequence of this is that the Wheeler-DeWitt equation should not be satisfied in quantum gravity. 
Classical states have a natural non-trivial time evolution since they are not eigenstates of the Hamiltonian. 
A general feature of the unconstrained quantum theory of gravity is the prediction of a pressureless dark matter component of either sign energy density in the classical universe which may lead to novel phenomenology.

\end{abstract}

\section{Introduction}
Quantum Mechanics and General Relativity are the two great pillars of modern physics. Together, they describe phenomena ranging from sub-nuclear scales to the size of the universe, spanning nearly 45 orders of magnitude in energy and time scales. While the ultra-violet nature of gravity is unknown, there is no doubt that its infrared behavior is governed by General Relativity. It is thus important to quantize classical General Relativity {\it i.e.} develop a framework where the gravitational degrees of freedom evolve quantum mechanically as opposed to their evolution via the equations of motion of classical General Relativity. This is important both to ensure the basic consistency of these two theories with each other and to identify potentially new dynamics that may be present in the quantum theory. 

This program has been successfully implemented in the perturbative regime where the phenomena of interest are restricted to the quantum mechanics of gravitons {\it i.e.} fluctuations around a classical background \cite{Donoghue:1994dn}. But, it is important to understand the quantum mechanical behavior of gravitation in the non-linear regime where the gravitational dynamics cannot be described solely as fluctuations around a classical background. This non-linear regime describes the behavior of gravitation around black holes and the cosmos. Unfortunately, the application of quantum mechanical principles to derive the correct quantum dynamics of non-linear gravitational systems has been plagued by a number of confusions.

 These confusions arise since gravitation is a gauge theory. Canonical quantization requires the identification of the physical degrees of freedom and a suitable elimination of the gauge degrees of freedom \cite{Dirac}. The direct application of the procedure to quantize electromagnetism on gravitation results in a trivial Hamiltonian \cite{Dirac, PhysRev.160.1113}. Since this procedure is unable to eliminate the gauge degrees of freedom without trivializing the theory, it was suggested that the gauge degree of freedom can be eliminated by restricting the Hilbert space of the theory to only include the energy eigenstates of the Hamiltonian. If this were to be true, the time evolution of physical states would be trivial. While some measure of time evolution can be recovered in asymptotically flat space-times, this procedure does not work for cosmological space-times. This confusion is also tied to the definition of time.  The Hamiltonian describes time evolution. In General Relativity, due to general covariance, there is no universal notion of time. One thus needs to recognize that time evolution should be identified as the relative evolution between the gravitational states and other states in the theory, with one of the states acting as a clock to keep track of the relative evolution. While this point is well known, the insistence that the universe is in an energy eigenstate implies that the dynamical world can be explained only if the universe was placed in an enormous superposition over all time \cite{Page:1983uc}. Such a state is not normalizable. Moreover,  the clock state needs to be integrated over all time in order to create this state. But, it is difficult to see how such a clock can be constructed to monotonically track relative evolution over all time using systems with bounded  Hamiltonians with quantum states that span a finite range of energies \cite{Unruh:1983ms}.  

In light of the above confusions, there has been considerable interest in developing a path integral approach to quantizing non-linear gravitational states \cite{Halliwell:1988wc}. This path integral formalism needs to handle the gauge redundancy in the system. Unfortunately, this problem has not been properly handled in existing work. Current approaches to this problem are focused on maintaining all of the classical equations of motion of General Relativity. This prevents these approaches from performing a sensible gauge fixing procedure. Without gauge fixing, the procedure yields divergent answers. 

The intent of our paper is to clarify these confusions and present a path integral approach to quantizing General Relativity. We focus our attention on the simplest scenario - namely mini-superspace quantum cosmological models where we only consider homogeneous  cosmological evolution \cite{Halliwell:1988wc}. We show that a suitably gauge fixed path integral yields finite transition amplitudes. We find that the gauge fixing procedure is necessary not only to get finite transition amplitudes but also to define the notion of co-ordinate time. The general covariance of the theory implies that this procedure maintains physical correlations between gravitational states and the matter states in the universe.  As a virtue of the path integral procedure, via the Schwinger-Dyson procedure, we show that the dynamical equations of motion of Einstein are automatically obeyed at the level of expectation values of the field operators. Our analysis shows that the Hamiltonian constraint equation that emerges in classical General Relativity is not a requirement of the quantum dynamics. But, this is not a problem since these classical constraints only need to be obeyed by the classical ({\it i.e.} coherent) states of the theory. We construct such states and a corresponding Hamiltonian that can be used to describe non-trivial time evolution of this gravitational system. This procedure also clarifies the role of the classical constraint equation in the dynamics of the quantum system. 

The rest of this paper is organized as follows. In section \ref{sec:quantization}, we describe the general principles of quantizing a theory using the path integral,  the construction of a Hamiltonian and classical coherent states consistent with this path integral. In section \ref{sec:QuantumCosmology}, we apply this path integral approach to quantizing mini-superspace cosmology. Using this formalism, we construct the corresponding Hamiltonian in section \ref{sec:Hamiltonian}. In section \ref{sec:example}, we apply these principles to an explicit model of a cosmology with a rolling scalar field. We then conclude in section \ref{sec:conclusions}. 

\section{Quantization}
\label{sec:quantization}

Given a physical system whose classical dynamics are known, how does one derive the appropriate quantum mechanical evolution? Since the classical dynamics are a subset of the overall quantum dynamics, it is not possible to derive the quantum mechanical evolution solely from the classical equations. The process of quantization requires us to make certain axiomatic assumptions. For example, in the canonical quantization procedure, this requires making assumptions about commutation (or anti-commutation) relations between fields and their conjugate momenta. We then assume that the time evolution of quantum states is governed by the Schrodinger equation, with a Hamiltonian constructed from the field operators. Ultimately, the only logical requirement of this procedure is that the time evolution of a subset of quantum states that correspond directly to classical states should automatically satisfy the classical equations of motion whenever such states obey the quantum equations of motion. In addition, one might also expect that the classical equations of motion hold in some ``average'' sense since the average behavior of a large number of quantum systems should resemble classical dynamics. 

We review how these issues are addressed in the path integral formulation of quantum mechanics. Given a classical Lagrangian $\mathcal{L}\left(\phi, \partial_{\mu} \phi\right)$ for some scalar field $\phi$, instead of the Schrodinger equation, the path integral formulation makes an axiomatic assumption about the time evolution of field basis states. That is, given the field operator $\phi\left(x\right)$, we consider its eigenstates $|\phi_e\rangle$ {\it i.e. } $\phi\left(x\right) |\phi_{e}\rangle = \phi_{e}|\phi_{e}\rangle$. The time evolution of these states is defined to be: 

\begin{equation}
    \langle \phi_f | T\left(t_2; t_1\right)|\phi_i\rangle = \int ^{\phi\left(t_2\right) = \phi_f}_{\phi\left(t_1\right) = \phi_i} D\phi \, e^{i \int^{t = t_{2}}_{t= t_1} d^4 x \, \mathcal{L}\left(\phi, \partial_{\mu} \phi \right)}
    \label{Eqn:PathIntegralDef}
\end{equation}

This defines the transition matrix element of the time evolution operator of an initial basis state $|\phi_i\rangle$ at time $t_1$  with the basis state $|\phi_f\rangle$ at time $t_2$. This axiomatic assumption leads to an immediate consequence. The path integral \eqref{Eqn:PathIntegralDef} should not change when the variable of integration $\phi$ is shifted to $\phi + \delta \phi$. Demanding the invariance of the path integral under variable redefintion, we obtain the so called Schwinger-Dyson relations which yield the analogs of the field equations. In the Heisenberg picture, this yields the equations\footnote{One can also construct analogous equations in the Schrodinger picture where now the derivatives in the field equations will act on the expectation values of the fields}:  
\begin{equation}
    \langle \Psi | \frac{\delta S}{\delta \phi\left(t\right)} | \Psi \rangle = 0 
    \label{SchwingerGeneral}
\end{equation}
where $S$ is the classical action $S = \int_{t_1}^{t_2} d^4x \mathcal{L}\left(\phi, \partial_{\mu}\phi\right)  $. This procedure is the analog of Ehrenfest's theorem in quantum field theory and it shows that the  classical equations are obeyed at the level of expectation values of normal ordered operators $\frac{\delta S}{\delta \phi}$. In proving the above, we assume that variations vanish at the boundaries. We also confess to an abuse of notation. Throughout this paper, we will discuss objects such as the expectation value described in \eqref{SchwingerGeneral}. In such expressions, objects such as $\frac{\delta S}{\delta \phi}$ are normal ordered operators and not classical fields.  

An important subtlety rises in the case of gauge theories. In such theories, the path integral \eqref{Eqn:PathIntegralDef} yields a finite answer only after gauge fixing. Thus the Lagrangian in \eqref{Eqn:PathIntegralDef} is the gauge fixed Lagrangian $\mathcal{L}_{F}$. The Schwinger-Dyson procedure is still valid but the identity  \eqref{SchwingerGeneral} it generates will involve the gauge fixed action $S_{F}$ as opposed to the classical action $S$. This identity can be different from the classical relation. There is nothing inconsistent about this since the quantum relations supersede classical equations. All that can be required is that for classical states, these relations reproduce the classical equations of motion. 

What are these classical states? To understand these, it is useful to switch from the path integral to the Hamiltonian formulation which more readily deals with states and their evolution. Before describing classical states, let us first address how one might obtain the corresponding Hamiltonian $H$ from the path integral. The path integral is defined in terms of the classical Lagrangian $\mathcal{L}\left(\phi, \partial_{\mu}\phi\right)$ where the fields and their derivatives are commuting classical fields. The Hamiltonian though is a function of  non-commuting field operators and their conjugate momenta. The procedure for computing the classical Hamiltonian from the classical Lagrangian thus yields a problem of ordering due to the non-commuting nature of the field operators. 

For a bosonic theory, we can resolve the issue of defining classical states and identifying the ordering of the Hamiltonian by making the following observation. We want the quantum theory to retain the symmetries of the classical theory. That is, we want the symmetries of the classical Hamiltonian to also be symmetries of the quantum Hamiltonian. To accomplish this, express the field and its conjugate momentum in terms of canonical creation and annihilation operators. Construct coherent states $|\Psi_c\rangle$  of these creation and annihilation operators - these are the classical states of this theory. Coherent states have the property that for any operator $O$, specification of the diagonal matrix elements $\langle \Psi_c | O | \Psi_c \rangle$ completely specifies the operator (for example, see \cite{ReedCoherent}). Thus, the Hamiltonian $H$ is completely determined by the expectation values $\langle \Psi_c | H | \Psi_c\rangle$. For coherent states $|\Psi_c\rangle$ one can show  \cite{Klauder:2021vld, Zhang:1999is}:

\begin{equation}
    \langle \Psi_c | :H: |\Psi_c\rangle = H_{c}\left( \langle \Psi_c| \Pi| \Psi_c\rangle,  \langle \Psi_c| \phi| \Psi_c\rangle\right)
    \label{DefHamiltonian}
\end{equation}

Here $:H:$ and $H_C$ are, respectively, the normal ordered and classical Hamiltonian obtained from the classical Lagranian $\mathcal{L}$. That is, the expectation value of the normal ordered Hamiltonian $:H:$ in a coherent state $|\Psi_c\rangle$ is equal to the classical Hamiltonian evaluated on the coherent state expectation values of the field and its conjugate momenta. From this construction, it is clear that  time evolution generated by $:H:$ preserves the symmetries of the corresponding classical theory. Thus, as long as the quantum dyanmics is generated by $:H:$, the symmetries of the classical theory are also maintained at the quantum level.  

The above statements about coherent states are likely more familiar to the reader in the context of field theories where the kinetic terms of the Hamiltonian are those of a free theory. But, this construction and result also applies for general field theories where the kinetic terms of the Hamiltonian are non-quadratic. To see this, suppose we have a Hamiltonian with non-quadratic kinetic terms in the field $\phi$ and its conjugate momentum $\Pi$. Now $\phi$ and $\Pi$ are simply operators on a Hilbert space. We can trivially imagine a free field theory constructed with $\phi$ and $\Pi$ and for this free field theory, we can define creation and annihilation operators using $\phi$ and $\Pi$ and construct a Fock space. This Fock space is an orthonormal basis on this Hilbert space. Now, consider the original non-linear theory. Even though the Fock space states constructed with the canonical creation and annihilation operators are not energy eigenstates of the non-linear theory, these Fock space states are still an orthonormal basis on the Hilbert space. When the canonical creation and annihilation operators act on these Fock space states, they will have the same algebra as the case of the free field theory since the operator algebra is independent of the underlying Hamiltonian. The construction and properties of coherent states simply relies on this operator algebra. Since this algebra also holds in the case of the non-linear Hamiltonian, we can use the creation and annihilation operators of the free field theory to define coherent states. The resultant statements about expectation values of normal ordered operators thus applies to these non-linear theories as well. The above construction will be central to our discussions about the Hamiltonian for General Relativity. We will use the requirement that the quantum evolution of coherent states reproduce the symmetries of classical General Relativity to resolve operator ordering problems in quantum General Relativity.

We have now described two different ways to time evolve a quantum state: the first using the path integral \eqref{Eqn:PathIntegralDef} and the second using the Schrodinger equation employing the normal ordered Hamiltonian \eqref{DefHamiltonian}. It can be shown that these two procedures yield the same time evolution using the conventional derivation of the path integral from the Schrodinger equation \cite{Zhang:1999is}. Henceforth, in this paper, when we refer to the Hamiltonian, we will always mean this normal ordered Hamiltonian and simply denote it as $H$ instead of $:H:$. 

In the following sections, we employ the above principles to obtain the time evolution of quantum cosmology. 

\section{Quantum Cosmology}
\label{sec:QuantumCosmology}

We describe the quantum mechanical evolution of a scalar field $\phi$ with potential $V\left(\phi\right)$ in General Relativity. We work in mini-superspace  where we restrict our attention to the homogeneous evolution of $\phi\left(t\right)$ and the metric $g_{\mu \nu}\left(t\right)$. In this limit, the metric $g_{\mu \nu}$ is: 

\begin{equation}
    g_{\mu \nu} \rightarrow ds^2 = - N\left(t\right)^2 dt^2 + a\left(t\right)^2 \left(dx^2 + dy^2 + dz^2\right)
    \label{Eqn:GRMetric}
\end{equation}

To describe quantum evolution, we need to know the physical degrees of freedom. In this system, these are $\phi\left(t\right)$ and $a\left(t\right)$, with $N\left(t\right)$ being a gauge degree of freedom. 

We formulate the quantum theory via the path integral formulation. In the path integral formulation what we wish to compute is the transition matrix element: 

\begin{equation}
    \langle \phi_f, a_f | T\left(t_2; t_1\right) | \phi_i, a_i\rangle 
\end{equation}
where $T\left(t_2; t_1\right)$ is the propagator that propagates an initial basis state $|\phi_i, a_i\rangle$ of the scalar field $\phi$ and the scale factor $a$ at time $t_1$ to a final basis state $\phi_f$ and $a_f$ at time $t_2$. 

There are two interrelated issues that arise when defining this path integral and they have a common resolution. The first issue is that since $N\left(t\right)$ is a gauge degree of freedom, the path integral can only be defined by gauge fixing $N\left(t\right)$. For this metric ansatz, a convenient gauge to use is the gauge  $\dot{N}\left(t\right) = 0$. The second issue arises from the definition of time. In General Relativity, time is simply a parameter - it can be redefined at will and thus one has to confront the question of what is meant by the definition of the propagator $ T\left(t_2; t_1\right)$ since the values $t_{1,2}$ can clearly be redefined. 

To resolve these issues, let us simply write down the standard formula for the gauge fixed path integral that yields the matrix elements of $T\left(t_2; t_1\right)$:

\begin{equation}
    \langle \phi_f, a_f | T\left(t_2; t_1\right) | \phi_i, a_i\rangle  = \int_{\phi\left(t_1\right) = \phi_1, a\left(t_1\right) = a_1, N\left(t_1\right) = N_1}^{\phi\left(t_2\right) = \phi_2, a\left(t_2\right) = a_2, N\left(t_2\right) = N_2}\,  DN \, D\phi \, Da  \, D\lambda \,  e^{i \int_{t = t_1}^{t = t_2} d^4 x  \left(\mathcal{L} - \lambda \dot{N} \right)} 
    \label{Eqn:QG}
\end{equation}
In the above, we should also include Gibbons-Hawking-York boundary terms to define the variational calculus for manifolds with a boundary. These will not be relevant for our discussion and thus we do not explicitly write them. In \eqref{Eqn:QG}, we have introduced a Lagrange multiplier $\lambda$ to enforce the gauge condition $\dot{N}=0$. The gauge condition also restricts the initial conditions of the function  $N\left(t_1\right) = N\left(t_2\right) = N_0$.  Setting $\dot{N} = 0$,  the integral over $N$ is trivial, with $N$ fixed at the boundary; in this case the condition  $\dot{N} = 0$ fully fixes the gauge.
We are thus reduced to evaluating: 
\begin{equation}
    \langle \phi_f, a_f | T\left(t_2; t_1\right) | \phi_i, a_i\rangle  = \int_{\phi\left(t_1\right) = \phi_1, a\left(t_1\right) = a_1, N\left(t_1\right) = N_0}^{\phi\left(t_2\right) = \phi_2, a\left(t_2\right) = a_2, N\left(t_2\right) = N_0}\,  D\phi \, Da  \, e^{i \int_{t = t_1}^{t = t_2} d^4 x  \mathcal{\tilde{L}}} 
    \label{Eqn:QGFixN}
\end{equation}

Here $\mathcal{\tilde{L}}$ is the gauge fixed Lagrangian after the path integral over  $N$ and $\lambda$ have been performed. This path integral is dominated by paths near the classical paths that are extrema of the action {\it i.e.} paths that satisfy the classical equations of motion: 

\begin{equation}
    \partial_t \left(\frac{\partial \mathcal{\tilde{L}}}{\partial \dot{\phi}}\right) - \frac{\partial \mathcal{\tilde{L}}}{\partial \phi} = 0
    \label{Eqn:scalarfield}
\end{equation}

\begin{equation}
    \partial_t \left(\frac{\partial \mathcal{\tilde{L}}}{\partial \dot{a}}\right) - \frac{\partial \mathcal{\tilde{L}}}{\partial a} = 0
    \label{Eqn:FRW2}
\end{equation}
It is easily checked that \eqref{Eqn:scalarfield} produces the usual second order differential equation for a scalar field $\phi$ while \eqref{Eqn:FRW2} yields the second (or spatial, $G_{ii} = -8 \pi G T_{ii}$) Friedman equation {\it i.e.} a second order differential equation for $a\left(t\right)$. Since both of these equations are second order differential equations, given the boundary conditions $\left(\phi\left(t_1\right) = \phi_i, a\left(t_1\right) = a_i \right)$ and $\left(\phi\left(t_2\right) = \phi_f, a\left(t_2\right) = a_f \right)$, we can find a classical solution. It can be shown that the path integral \eqref{Eqn:QGFixN} is finite \cite{Halliwell:1988wc} and thus defines a sensible propagator  $T\left(t_2; t_1\right)$.

What would happen if we picked a different value for the gauge parameter $N$ {\it i.e.} set $N_0 \rightarrow \tilde{N_0}$ in the path integral \eqref{Eqn:QGFixN}? From the metric \eqref{Eqn:GRMetric}, a redefinition $N_0 \rightarrow \tilde{N_0}$ is a redefinition of the time co-ordinate. Now  consider a path $\left(\phi\left(t\right), a\left(t\right) \right)$ from $\left(\phi\left(t_1\right) = \phi_i, a\left(t_1\right) = a_i \right)$ to $\left(\phi\left(t_2\right) = \phi_f, a\left(t_2\right) = a_f \right)$ in the $a-\phi$ plane.  When we set $N_0 \rightarrow \tilde{N_0}$, this is simply a reparameterization $\left(\phi\left(\tilde{t}\right), a\left(\tilde{t}\right) \right)$ of the same physical path in the $a-\phi$ plane where we get $\tilde{t}$ from the reparameterization determined by $N_0 \rightarrow \tilde{N_0}$. Since the action of General Relativity is invariant under time re-parameterization, we then have: 

\begin{equation}
    \langle \phi_f, a_f| \tilde{T}\left(\tilde{t_2}; \tilde{t_1}\right) | \phi_i, a_i\rangle = \langle \phi_f, a_f| T\left(t_2; t_1\right) | \phi_i, a_i\rangle  
    \label{Eqn:RelateElements}
\end{equation}

That is, the transition matrix element of the propagator between the physical states $|\phi_i, a_i\rangle$ and $|\phi_f, a_f\rangle$ are the same as long as the time co-ordinates in the propagator are suitably rescaled. Thus, the time evolution of the basis state $|\phi_i, a_i\rangle$ is: 

\begin{equation}
    |\phi_i, a_i\rangle \rightarrow \sum_{\phi_f, a_f} c_{\phi_f, a_f}\left(t_2, t_1\right) |\phi_f, a_f\rangle = \sum_{\phi_f, a_f} \tilde{c}_{\phi_f, a_f}\left(\tilde{t_2}, \tilde{t_1}\right) |\phi_f, a_f\rangle
    \label{Eqn:universesuperposition}
\end{equation}
where the coefficients $c_{\phi_f, a_f}\left(t_2, t_1\right)$ and $\tilde{c}_{\phi_f, a_f}\left(\tilde{t_2}, \tilde{t_1}\right)$ are the transition matrix elements \eqref{Eqn:RelateElements}.  Given this time evolution of the basis states, we see that the initial quantum state $|\Psi\left(t_1\right)\rangle = |\tilde{\Psi}\left(\tilde{t_1}\right)\rangle = |\Sigma\rangle$ becomes the final state $|\Psi\left(t_2\right)\rangle = |\tilde{\Psi}\left(\tilde{t_2}\right)\rangle = |\Omega\rangle$.

This  clarifies the issue of time that has plagued discussions about quantum cosmology. Time evolution is the statement that a given initial quantum state $|\Sigma\rangle$ evolves to a state $|\Omega\rangle$. This evolution takes a certain co-ordinate time. But, co-ordinate time by itself is meaningless in General Relativity. The co-ordinate time can be changed simply by picking a different value of $N_0$. What does this time evolution in co-ordinate time mean?  In the above, we have shown the following. Let us pick some value of $N_0$. This defines a co-ordinate time $t$. We can pick an initial quantum state $|\Psi\left(t_1\right)\rangle = |\Sigma\rangle$. Using the path integral, we can now compute the final state $|\Psi\left(t_2\right)\rangle = |\Omega\rangle$. Instead of $N_0$, suppose we pick a different value $\tilde{N_0}$. This picks a different definition of the co-ordinate time $\tilde{t}$. As long as we rescale the times suitably $t_1 \rightarrow \tilde{t_1}$ and $t_2 \rightarrow \tilde{t_2}$, we have the same physical effect: $|\Psi\left(t_1\right)\rangle = |\tilde{\Psi}\left(\tilde{t_1}\right)\rangle = |\Sigma\rangle \rightarrow |\Psi\left(t_2\right)\rangle = |\tilde{\Psi}\left(\tilde{t_2}\right)\rangle = |\Omega\rangle$. We thus learn that no matter what value of $N_0$ is chosen, the physical evolution of the quantum states in the $a-\phi$ Hilbert space is the same - the time co-ordinate is just a convenient parameterization of this evolution.\footnote{In a deeper sense, time evolution should be regarded as the relative evolution between different quantum states. In general, one could try to formulate time evolution without co-ordinate time by writing the laws of nature in terms of partial derivatives of various fields. Since the motion is along a complicated Hilbert space, such an equation is likely to be complex. It is easier to describe this motion using a simple parameterization, namely, time.}

It is important to note that even though the co-ordinate time label that is used to define time evolution can be changed by picking a different value of $N_0$, for any given $N_0$, the co-ordinate label is extremely meaningful. Time evolution, by its very definition, is the notion that an initial quantum state $|\Sigma\rangle$ changes over a certain co-ordinate time to become a different state $|\Omega\rangle$. We see that the gauge fixing that is necessary to perform the path integral ({\it i.e.} picking a value of $N_0$) fixes the meaning of the co-ordinate time that appears in the time evolution operator. Without a definition of a co-ordinate time, this time evolution operator is meaningless. 

This is the key difference between our treatment of the path integral and the conventional approach. In the conventional approach, once the finite path integral over the dynamical degrees of freedom $\phi$ and $a$ is performed, an integral is performed over all possible values of $N_0$. This makes the path integral independent of the gauge fixing parameter $N_0$ while yielding a divergent answer. But, this is hardly surprising - $N$ is a gauge degree of freedom and an integration over a gauge degree of freedom is expected to be divergent. This integral over $N$ should not be performed.  In physical terms, when a sum over all $N$ is performed, the question that is being asked is: if the quantum state is initially in the basis state $|\phi_i, a_i\rangle$, what is the transition matrix element for the state to become the basis state $|\phi_f, a_f\rangle$ over {\bf all} time? It is unsurprising that this answer is ill defined and divergent - the basis states have a wide spread of conjugate momenta and thus when integrated over all time, they can transition into each other in infinitely many ways. The physical meaning of time evolution is not the amplitude for transitioning from an initial quantum state to some final quantum state over the entire history of the universe. Rather, the physical meaning of time evolution is the actual path taken by the quantum state in Hilbert space to go from an initial state to some final state. Time is a parameterization of this path.  It is thus fully appropriate for the path integral, which in fact describes this evolution in a particular choice of time co-ordinate, to depend on this choice via the gauge parameter $N_0$. We have seen that despite this dependence, the physical outcomes are independent of the choice of $N_0$. 

Another major difference between our approach and the conventional approach is that the procedure outlined above does not involve the constraint equation obtained from the first (or temporal,\footnote{Note we have used Weinberg's convention in defining the Riemann tensor, which we adopt for this paper.} $G_{tt} = - 8 \pi G T_{tt}$) Friedman equation.\footnote{The analogous approach in electromagnetism would be not to  invoke the requirement that physical states obey Gauss' law, $\nabla.E\,|\Psi\rangle_{\text{physical}}=0$. The unconstrained theory has a classical non-zero background charge density that is conserved because $[H,\nabla.E]=0$. We will see below that the counterpart of this background in the gravity theory is a pressureless dark matter component.}  In the conventional approach, since the path integral over $N\left(t\right)$ is retained, the constraint equation is required to hold, resulting in the Wheeler-DeWitt equation. In our procedure, the path integral over $N$ is trivial due to the gauge choice $\dot{N}\left(t\right) = 0$ and the fixed boundary conditions (or) choice of time coordinate. The classical paths from $\left(\phi\left(t_1\right) = \phi_i, a\left(t_1\right) = a_i \right)$ to $\left(\phi\left(t_2\right) = \phi_f, a\left(t_2\right) = a_f \right)$ are not required to satisfy this constraint equation and thus we do not require the Wheeler-DeWitt equation to hold. In classical General Relativity, the constraint equation restricts the initial conditions for the second order dynamical equation for the scale factor -- the second/spatial Friedman equation. Classically, this implies that in this system, we are only allowed to specify $\phi\left(t_1\right) = \phi_1, \phi\left(t_2\right) = \phi_2$, and $a\left(t_1\right) = a_i$.  Due to the constraint equation, there is no freedom to specify $a\left(t_2\right) = a_f$ and the scale factor $a\left(t_2\right)$ is dynamically determined. In quantum evolution, due to the absence of this constraint, one in general gets a spread of metric states $a\left(t_2\right)$ correlated with the scalar field state $\phi\left(t_2\right)$ as opposed to the unique correlation between $\phi\left(t_2\right)$ and $a\left(t_2\right)$ in the classical case. 

Let us now comment on the connection between this quantization procedure and the classical field equations. The path integral, as a virtue of the Schwinger-Dyson procedure discussed above, guarantees the following identities \footnote{We write these in the Heisenberg picture for notational simplicity. In the Schrodinger picture, the derivatives act on expectation values.}: 

\begin{eqnarray}
  \langle \Psi |  \left( \partial_t \left(\frac{\partial \mathcal{L}}{\partial \dot{\phi}}\right) - \frac{\partial \mathcal{L}}{\partial \phi}\right)|\Psi\rangle = 0 \nonumber \\ 
  \langle \Psi | \left(\partial_t \left(\frac{\partial \mathcal{L}}{\partial \dot{a}}\right) - \frac{\partial \mathcal{L}}{\partial a}\right) |\Psi\rangle  = 0
  \label{ClassicalEinstein}
\end{eqnarray}

The equations \eqref{ClassicalEinstein}  are the analogs of Ehrenfest's equations that correspond to the second order field equations for $\phi$ and $a$. However, when we perform the variation $\delta N$ on $N$, we get the equation: 

\begin{equation}
    \langle \Psi | \left( \frac{d\lambda}{dt}  - \frac{\partial \mathcal{L}}{\partial N}\right) |\Psi \rangle = 0
\label{Eqn:FirstEinstein}
\end{equation}
This is not a constraint on  $\langle \Psi | \frac{\partial \mathcal{L}}{\partial N} |\Psi\rangle $ - rather, it describes the evolution of the Lagrange multiplier $\lambda$ in the path integral. There is thus no physics associated with this equation. We thus see that the Wheeler-DeWitt equation does not follow from the path integral and is thus not an actual requirement of the quantum dynamics (see \cite{Cotler:2020lxj,Cotler:2021cqa } for similar phenomena in the Euclidean path integral).

The fact that \eqref{Eqn:FirstEinstein} is not of the form $\langle \Psi | \frac{\partial \mathcal{\tilde{L}}}{\partial N} |\Psi\rangle = 0$, the quantum analog of the classical first order constraint  $\frac{\partial \mathcal{\tilde{L}}}{\partial N} = 0$ that arises from classical General Relativity, should not bother us. In a typical quantum theory, since the Schrodinger equation is a first order differential equation, the quantum state of the physical degrees of freedom can be freely specified as initial conditions. The Schrodinger equation then yields the dynamical evolution of this initial state. The imposition of the constraint $\langle \Psi | \frac{\partial \mathcal{\tilde{L}}}{\partial N} |\Psi\rangle = 0$  is a restriction on the allowed initial state. But, there do not appear to be any inconsistencies in the quantum theory by relaxing this constraint since gauge fixing the gauge degree of freedom $N$ can be done independent of this constraint. To further understand the analog of the classical first order constraint equation of General Relativity it is useful construct the Hamiltonian described by the path integral \eqref{Eqn:QGFixN}, which we do in the next section.

Let us now discuss the invariance of the quantum dynamics under a more general gauge than the $\dot{N} = 0$ gauge. The gauge symmetry of General Relativity is general covariance {\it i.e.} the space-time manifold can be parameterized by any choice of coordinates and the underlying quantum dynamics should be invariant under such a reparameterization. In mini-superspace, this translates into the statement that we can take $t \rightarrow N\left(t\right)$  for arbitrary $N\left(t\right)$ and the quantum dynamics should still be invariant. Importantly, once we choose these coordinates, we have to perform the time evolution of the full quantum state using these coordinates. That is, if we have a quantum superposition, under a diffeomorphism, the time coordinate in the entire superposition is changed by the {\bf same} function $N\left(t\right)$. The diffeomorphism, being a function on the manifold, does not allow us to choose different time parameterizations for different parts of a superposition. This has implications for certain conventionally used gauge choices in classical General Relativity that cannot be blindly applied at the quantum level. 

In classical General Relativity, we are only interested in describing the behavior of a single metric and thus it is convenient to choose various gauges that depend on the metric itself - for {\it e.g.} one may take $N\left(t\right) = f\left(a\left(t\right)\right)$. In quantum mechanics, we are required to describe the evolution of superpositions of metrics - indeed the path integral itself is a computation that specifies quantum evolution as a sum over many such superpositions. It is tempting to think that one could take the gauge $N\left(t\right) = f\left(a\left(t\right)\right)$ in the path integral and integrate over the dynamical degree of freedom $a\left(t\right)$ while performing this path integral and demand that the final result be reparameterizaton invariant. But, this is incorrect. Such a procedure specifies a different parameterization of the time coordinate for each part of the quantum superposition and it will not give time reparameterization invariant results. 

This is true even classically: one can choose a simple gauge $N\left(t\right) = 1$ and pick two different sets of initial values of fields and their cojugate momenta $\left(\phi_{1i},\Pi_{1i},a_{1i},\Pi^{a}_{1i}\right)$ and $\left(\phi_{2i},\Pi_{2i},a_{2i},\Pi^{a}_{2i}\right)$ on an initial time slice in the manifold. Time evolve this state to a final time where we have $\left(\phi_{1i},\Pi_{1i},a_{1i},\Pi^{a}_{1i}\right) \rightarrow \left(\phi_{1f},\Pi_{1f},a_{1f},\Pi^{a}_{1f}\right)$ and $\left(\phi_{2i},\Pi_{2i},a_{2i},\Pi^{a}_{2i}\right) \rightarrow \left(\phi_{2f},\Pi_{2f},a_{2f},\Pi^{a}_{2f}\right)$. Let us now pick a gauge of the form $N\left(t\right) = f\left(a\left(t\right)\right)$ and perform the same time evolution by using the respective metrics $a_{1,2}$ in this lapse function.

One can check that classically when $\left(\phi_{1i},\Pi_{1i},a_{1i},\Pi^{a}_{1i}\right) \rightarrow \left(\phi_{1f},\Pi_{1f},a_{1f},\Pi^{a}_{1f}\right)$, we do {\bf not} have $ (\phi_{2i},\Pi_{2i}, \allowbreak a_{2i}, \Pi^{a}_{2i}) \rightarrow \left(\phi_{2f},\Pi_{2f},a_{2f},\Pi^{a}_{2f}\right)$ {\it i.e.} the relative evolution is not gauge invariant. But, this is not a surprise since the parameterization of time is being picked differently for each metric. What is true however is that we are free to choose a gauge of the form $N\left(t\right) = f\left(a_1\left(t\right)\right)$ {\bf or} $N\left(t\right) = f\left(a_2\left(t\right)\right)$ and use the same gauge to perform both of these evolutions. In this case, we will get gauge invariant results. 

In other words, the function $N\left(t\right)$ can be any arbitrary function of time (as demanded by general covariance) but it is the same function of time that has to be applied across the entire quantum superposition. One cannot have the dynamical degree of freedom appearing in $N\left(t\right)$ with this degree of freedom being integrated over in the path integral.  It is straightforward to demonstrate gauge invariance of the physical evolution of the path integral parameterized by  such an arbitrary function of time (that applies across the full quantum state) by suitably rescaling the time parameters as articulated in the discussion of the $\dot{N}=0$ gauges.

\section{Hamiltonian Construction}
\label{sec:Hamiltonian}

This path integral procedure  sheds light on how one may canonically quantize gravity. When Dirac \cite{Dirac} tried to canonically quantize gravity, he ran into the following issue: in mini-superspace, the physical degrees of freedom are $a$ and $\phi$ while $N$ is a gauge degree of freedom. To quantize the theory, it would seem like one has to decide what the $N$ operator is supposed to be. Following the procedure in electromagnetism, Dirac first tried to solve for $N$ by requiring the temporal Einstein equation $G_{00} + 8 \pi G T_{00} = 0$  to hold at the operator level. But, since $H_N = -N \left(G_{00} + 8 \pi G T_{00}\right)$, if $\left(G_{00} + 8 \pi G T_{00}\right) = 0$, we also have $H_N = 0$ {\it i.e.} the Hamiltonian vanishes identically and the theory is trivial\footnote{The situation is different in electromagnetism where $(\nabla\cdot E)$ can be set to zero at the operator level and one can still obtain a non trivial Hamiltonian.}. Instead, Dirac imposed the requirement that the physical states of the theory satisfy the temporal Einstein equation $\left(G_{00} + 8 \pi G T_{00}\right) |\Psi\rangle = 0$, resulting in the Wheeler deWitt equation  $H_N |\Psi\rangle = 0$. 

While this procedure does not determine $N$, this restriction on the Hilbert space makes the quantum theory independent of $N$. This is similar to the procedure followed in gauge theories such as electromagnetism. In certain covariant gauges like the Lorenz gauge, the gauge redundancy is not fully removed and the Hamiltonian depends upon gauge degrees of freedom. But, by requiring physical states to obey certain constraint equations, one can prevent the gauge degrees of freedom from being physically relevant. It is thus tempting to attempt the same procedure in General Relativity since $H_N$ depends on $N$. While the Hamiltonian $H$ of this theory is now non-trivial, if $H_N |\Psi\rangle = 0$, then every physical state $|\Psi\rangle$ is required to be an energy eigenstate of $H$  and thus the state has no evolutionary dynamics. Note that in this case, all the physical states have to have the same energy eigenvalue since if they did not, we could take a superposition of them which would then not be an energy eigenstate. This is the famous ``problem of time'' of quantum cosmology \cite{PhysRev.160.1113}. 

Let us see how the path integral formulation can be used to canonically quantize gravity. In this formulation, the meaning of $N$ is clear - $N$ defines co-ordinate time. Thus, one should not be ``solving'' for it - one simply chooses it. That is, $N$ is not an operator, it is simply the number $N_0$ that was picked to define co-ordinate time in the path integral.  One can now take the Lagrangian $\mathcal{\tilde{L}}$ obtained by gauge fixing and construct the canonical normal ordered Hamiltonian $H_N$ from it. Quantum states will evolve as per the Schrodinger equation:

\begin{equation}
    i \frac{\partial |\Psi\rangle}{\partial t_N} = H_{N} |\Psi\rangle
    \label{Eqn:QuantumSchrod}
\end{equation}
reproducing the dynamics of the path integral obtained from the gauge fixed Lagranian. Now, different values of $N$ will yield different Hamiltonians $H_N$ - but this is simply the statement that choosing different values of $N$ corresponds to choosing different time co-ordinates $t_N$. Even though these Hamiltonians are different, just as in the case of the path integral, the evolution will preserve  correlations between quantum states {\it i.e. } the physical meaning of ``time evolution''. We work this out explicitly in section \ref{sec:example}. We see that the nature of the gauge symmetry in General Relativity is different from that of electromagnetism. In electromagnetism, one could solve for the gauge operators in terms of other physical operators. But, this is not true in General Relativity - $N$ picks the time co-ordinate and the time reparameterization invariance of the theory guarantees that the underlying physics is gauge invariant.

Importantly, this quantization procedure does not require the Wheeler deWitt equation to hold {\it i.e.} physical states $|\Psi\rangle$ of the theory are not required to be annihilated by the Hamiltonian of General Relativity. With  $H_N |\Psi\rangle \neq 0$,  the quantum theory does not automatically satisfy the temporal Einstein  equation  $\frac{\partial \mathcal{\tilde{L}}}{\partial N} |\Psi\rangle = 0$ (where $\frac{\partial \mathcal{\tilde{L}}}{\partial N}$  is now an operator).  But, the belief that the quantum theory should exactly satisfy the classical Einstein equation is incorrect.  The quantum theory supersedes the classical equations of motion - instead of restricting the allowed quantum states, what we must do is to find the analog of the classical equations of motion that automatically arise from the quantum evolution. 

Let us now discuss the analog of the first order constraint. Notice that any quantum state $|\Psi\rangle$ that satisfies \eqref{Eqn:QuantumSchrod} automatically satisfies: 

\begin{equation}
    \frac{d \langle \Psi | H_{N} |\Psi\rangle }{dt_N} = 0
    \label{Eqn:QuantumConstraint}
\end{equation}

Thus, if we are given an initial condition that satisfies the constraint $\langle \Psi | H_{N} |\Psi\rangle =  0$, subsequent time evolution will automatically obey the first order Einstein constraint $\langle \Psi | \frac{\partial \mathcal{\tilde{L}}}{\partial N}  |\Psi\rangle = 0$. 

With this understanding, let us now understand why the classical evolution of General Relativity appears to obey the first order constraint. For classical cosmology, we need states $|\Psi_c\rangle$ that are coherent states of the theory (and not energy eigenstates as conventionally demanded). The construction of these coherent states in this non-linear theory is performed using the canonical creation and annihilation operators of the corresponding fields, as discussed in section \ref{sec:quantization}. For these states, the expectation value $\langle \Psi | \frac{\partial \mathcal{\tilde{L}}}{\partial N} |\Psi\rangle$ reduces to the classical expression for $\frac{\partial \mathcal{\tilde{L}}}{\partial N}$ evaluated on the expectation values $\langle a\rangle$, $\langle \Pi_a \rangle$, $\langle \phi \rangle$ and $\langle \Pi\rangle$. If we choose a coherent state that obeys the Hamiltonian constraint $\langle \Psi_c | \frac{\delta \mathcal{\tilde{L}}}{\delta N} | \Psi_c\rangle = 0$, its subsequent time evolution will automatically obey the first order classical constraint of General Relativity.  This procedure thus sheds light on the reason why classical General Relativity is able to solve the initial value problem. Quantum evolution is described by first order differential equations. Thus, given any initial state one can time evolve it. This is unlike the over constrained classical equations of General Relativity. But, the quantum equations guarantee that if we are given an initial state that obeys the first order constraint of classical General Relativity, the evolution will continue to preserve it. This is precisely the initial value problem of General Relativity and it is trivially proven using quantum dynamics.

What if we picked a coherent state that did not obey this constraint? Relaxation of this condition would imply that there would be coherent states of the theory that satisfy the dynamical Einstein equations \eqref{ClassicalEinstein} with $\langle \Psi | \frac{\partial \mathcal{\tilde{L}}}{\partial N} |\Psi\rangle \neq 0$, leading to the existence of classical solutions that apparently violate the classical Einstein constraint. However, this does not seem to be an actual problem. When we have a coherent state with $\langle \Psi | \frac{\partial \mathcal{\tilde{L}}}{\partial N} |\Psi\rangle \neq 0$, this implies that the expectation value $\langle \Pi_a\rangle$ ({\it i.e.} the Hubble parameter) is not commensurate with the energy densities in the universe. But, the path integral allows us to time evolve this state. This time evolution is identical to that of a classical state with the same initial value $\langle \Pi_a\rangle$ and a new pressure-less ``dark matter'' component whose initial energy density is the amount necessary to enforce the constraint $\langle \Psi | \frac{\partial \mathcal{\tilde{L}}}{\partial N} |\Psi\rangle = 0$. This is because such a pressure-less dark matter component only affects the constraint equation but not the dynamical evolution equations for $a$ and $\phi$. In other words, the gravitational excitations that lead to $\langle \Psi | \frac{\partial \mathcal{\tilde{L}}}{\partial N} |\Psi\rangle \neq 0$ effectively act as an initial state with a component of dark matter.  

\section{Examples}
\label{sec:example}
Let us next see how to apply the above formalism in a cosmological model where the scalar field potential $V\left(\phi\right) = 0$. With the metric \eqref{Eqn:GRMetric}, the Lagrangian for this theory is $\mathcal{L} = \frac{M_{pl}^2}{2} \sqrt{-g} R + \sqrt{-g} \frac{g^{\mu \nu} \partial_{\mu}\phi \partial_{\nu}\phi}{2}$, yielding: 

\begin{equation}
    \mathcal{L} = 3 M_{pl}^2\frac{a\left(t\right) \dot{a}\left(t\right)^2}{N\left(t\right)} - \frac{a\left(t\right)^3 \dot{\phi}\left(t\right)^2}{2 N\left(t\right)}
\end{equation}

 To proceed, we perform the gauge fixing procedure described in section \ref{sec:QuantumCosmology}  by taking $N\left(t\right) = \tilde{N}$, a constant, yielding: 

\begin{equation}
    \mathcal{L_F} = \frac{\left(3 M_{pl}^2a\left(t\right) \dot{a}\left(t\right)^2 - \frac{a\left(t\right)^3 \dot{\phi}\left(t\right)^2}{2 }\right)}{\tilde{N}}
\end{equation}

With this gauge fixed Lagrangian, we can compute the conjugate momenta $\Pi_{a} = \frac{\partial \mathcal{L_F}}{\partial \dot{a}}$, $\Pi = \frac{\partial \mathcal{L_F}}{\partial \dot{\phi}}$ and obtain the Hamiltonian $H = \Pi_{a} \dot{a} + \Pi \dot{\phi} - \mathcal{L_F}$, yielding: 

\begin{equation}
    H_{\tilde{N}} =\tilde{N}\left( a^{-1}\frac{\Pi_{a}^2}{12 M_{pl}^2} - a^{-3}\frac{\Pi^2}{2}\right)
\end{equation}
upto ordering ambiguities that we will resolve below. The conjugate momentum $\Pi_a$ satisfies the commutation relation $[a, \Pi_a] = i$ and thus in the ``position'' basis of the $a$ operator, $\Pi_a = -i \frac{\partial}{\partial a} $.  Time evolution is then determined using the Schrodinger equation with this Hamiltonian: 

\begin{equation}
    i \frac{\partial |\Psi\rangle}{\partial t_{\tilde{N}}} =  \tilde{N}\left( a^{-1}\frac{\Pi_{a}^2}{12 M_{pl}^2} - a^{-3}\frac{\Pi^2}{2}\right) |\Psi\rangle
    \label{Eqn:QuantumSchrod2}
\end{equation}

When a different value of $\tilde{N}$ is picked, \eqref{Eqn:QuantumSchrod2} makes it clear that it is simply just a reparameterizaton of the time coordinate. Physical correlations between states are determined by the time independent operators $a, \Pi_a, \Pi$ and $\phi$ - these correlations will be maintained under reparameterization of the $t$ coordinate. 

Since $a$ and $\Pi_a$  do not commute, we can now see why there is a question about the order of the operators in \eqref{Eqn:QuantumSchrod2} - should the kinetic term for $a$ be $a^{-1}\Pi_a^2$ (as written) or some linear combination of $a^{-1}\Pi_a^2$,  $\Pi_a a^{-1} \Pi_a$ and $\Pi_a^2 a^{-1}$?  To answer this question, we resort to the physical requirement stated in section \ref{sec:quantization} - namely, we demand that the quantum evolution of coherent states  $|\phi\Pi,a\Pi_a\rangle$ preserves the symmetries of the classical theory whose classical field values are given by the expectation values $\langle a \rangle = \langle \phi\Pi,a\Pi_a | a | \phi\Pi,a\Pi_a\rangle$, $\langle \Pi_{a} \rangle = \langle \phi\Pi,a\Pi_a | \Pi_{a} | \phi\Pi,a\Pi_a\rangle$, $\langle \phi \rangle = \langle \phi\Pi,a\Pi_a | \phi | \phi\Pi,a\Pi_a\rangle$ and $\langle \Pi \rangle = \langle \phi\Pi,a\Pi_a | \Pi | \phi\Pi,a\Pi_a\rangle$. In the absence of operators of the form $a^{-1}$ and $a^{-3}$ in \eqref{Eqn:QuantumSchrod2}, this requirement results in normal ordering the Hamiltonian as discussed in section \ref{sec:quantization}. But, since \eqref{Eqn:QuantumSchrod2} contains these inverse operators, the procedure is as follows. The correct operator that appears in  \eqref{Eqn:QuantumSchrod2} in place of $a^{-1}\Pi_{a}^2$ is the operator $O$ for which the expectation value $\langle \phi\Pi,a\Pi_a|O|\phi\Pi,a\Pi_a
\rangle$ is equal to the classical value $\langle \Pi_a\rangle^2/\langle a \rangle$. This specifies the diagonal matrix elements of $O$ on any coherent state. Since the coherent states are an overcomplete basis, it can be shown (for {\it e.g.} see \cite{ReedCoherent}) that the specification of these diagonal matrix elements uniquely determines $O$. Thus, the operator  $O$ that appears in the Hamiltonian in place of $a^{-1}\Pi_a^2$ is uniquely determined. The same argument also applies for the operator that appears instead of $a^{-3} \Pi^2$.

\section{Conclusions}
\label{sec:conclusions}

In this paper, we have shown that the quantum cosmology of mini-superspace can be defined unambiguously with a gauge-fixed path integral. This procedure yields finite transition amplitudes and it can be used to construct a non-trivial Hamiltonian. Classical evolution is described by coherent states of this theory. Since these are not energy eigenstates, the time evolution of this state is non-trivial. These states do not obey the Wheeler-DeWitt equation.  That is, we have found that the typical Hamiltonian constraint that is imposed in classical General Relativity is not required in the quantum theory. When this condition is not imposed, the evolution of the universe can still be classical but with the additional gravitational excitations acting as a form of dark matter. 

Our analysis focused on mini-superspace in order to simplify calculations and identify the key physical principles. It would be interesting to extend this analysis to more general cosmologies where one allows for inhomogeneous excitations. This would require gauge fixing components of the metric that mix time and space. Presumably, this gauge fixing procedure results in the identification of the foliation of the space-time manifold. The gauge fixed path integral should still yield finite transition matrix elements and this path integral can then be used to obtain a suitable Hamiltonian. 

In linear quantum mechanics, due to decoherence, it is difficult to conceive of scenarios where a quantum description of gravitation is necessary to understand the dynamics of the universe. The primary purpose of our work is to demonstrate that non-linear gravitational phenomena can be consistently described by quantum mechanics. The recognition that the classical states that do not satisfy the Hamiltonian constraint can be treated as classical states that obey the constraint but with additional dark matter may yield new phenomenology. The theoretical formalism developed in this paper may be useful to understand stability issues in certain cosmologies where quantum tunneling could potentially be important \cite{Graham:2014pca, Graham:2011nb, Mithani:2011en, Mithani:2014toa}. This formalism also permits  a much wider class of quantum states that could describe the universe as opposed to the energy eigenstates that are allowed by the Wheeler-DeWitt equation. These may find use in describing the initial state of the universe emerging from the big bang singularity.  Further, we also see that the quantum dynamics of coherent states is fully captured by the appropriate classical equations of motion. Since these states are a basis, understanding their classical behavior is tantamount to describing the actual quantum evolution of these systems. 

The most phenomenologically interesting applications of this formalism are likely to be found in non-linear quantum mechanics where there can be macroscopic quantum mechanical phenomena that exist despite decoherence \cite{Kaplan:2021qpv}. These phenomena exist when the universe is in a macroscopic superposition, as expected in conventional inflationary cosmology. The time evolution of such a state cannot be described by classical General Relativity. The formalism developed in this paper can be applied to these non-linear quantum mechanical theories to obtain cosmological observables.

\section*{Acknowledgments}
We thank Raman Sundrum, Steffen Gielen, and Shreya Shukla for discussions. D.~E.~K and S.R.~are supported in part by the U.S.~National Science Foundation (NSF) under Grant No.~PHY-1818899.  T.~M. is supported by the World Premier International Research Center Initiative (WPI) MEXT, Japan, and by JSPS KAKENHI grants JP19H05810, JP20H01896, and JP20H00153. 
This work was supported by the U.S.~Department of Energy (DOE), Office of Science, National Quantum Information Science Research Centers, Superconducting Quantum Materials and Systems Center (SQMS) under contract No.~DE-AC02-07CH11359. 
S.R.~is also supported by the DOE under a QuantISED grant for MAGIS, and the Simons Investigator Award No.~827042.

\bibliographystyle{unsrt}
\bibliography{references}
\end{document}